\documentclass[prb,aps,twocolumn,floats,showpacs]{revtex4}

\usepackage{graphics}
\usepackage{amsmath}
\usepackage{dcolumn}
\usepackage{citesort}
\usepackage{epsfig}
\begin{document}

\title {{\rm\small\hfill (submitted to Phys. Rev. B)}\\
On the accuracy of first-principles lateral interactions:
Oxygen at Pd(100)}

\author{Yongsheng Zhang}
\affiliation{Fritz-Haber-Institut der Max-Planck-Gesellschaft, Faradayweg
4-6, D-14195 Berlin, Germany}

\author{Volker Blum}
\affiliation{Fritz-Haber-Institut der Max-Planck-Gesellschaft, Faradayweg
4-6, D-14195 Berlin, Germany}

\author{Karsten Reuter}
\affiliation{Fritz-Haber-Institut der Max-Planck-Gesellschaft, Faradayweg
4-6, D-14195 Berlin, Germany}

\received{19th January 2007}

\begin{abstract}
We employ a first-principles lattice-gas Hamiltonian (LGH) approach to determine the lateral interactions between O atoms adsorbed on the Pd(100) surface. With these interactions we obtain an ordering behavior at low coverage that is in quantitative agreement with experimental data. Uncertainties in the approach arise from the finite LGH expansion and from the approximate exchange-correlation (xc) functional underlying the employed density-functional theory energetics. We carefully scrutinize these uncertainties and conclude that they primarily affect the on-site energy, which rationalizes the agreement with the experimental critical temperatures for the order-disorder transition. We also investigate the validity of the frequently applied assumption that the ordering energies can be represented by a sum of pair terms. Restricting our LGH expansion to just pairwise lateral interactions, we find that this results in effective interactions which contain spurious contributions that are of equal size, if not larger than any of the uncertainties e.g. due to the approximate xc functional.
\end{abstract}



\maketitle

\section{Introduction}

Lateral interactions between species adsorbed at solid surfaces are crucial microscopic quantities that have been the target of surface science studies for a long time.\cite{masel96} These interactions govern both the equilibrium, as well as the non-equilibrium ordering behavior of the adsorbates, and thereby critically influence the surface function and properties in important applications like heterogeneous catalysis. Traditionally, considerable efforts have been devoted to determine lateral interactions empirically from experimental data, e.g. from temperature programmed desorption or low energy electron diffraction (LEED) measurements. In order to simplify the inherently indirect determination from sparse experimental data, the assumption of exclusively pairwise interactions between the adsorbed species has often been applied. In recent years, algorithmic advances and increased computational power have made it possible to determine the lateral interactions alternatively from first-principles. Most notably, these are approaches that parameterize lattice-gas Hamiltonians (LGHs) with density-functional theory (DFT) energetics, also called cluster expansions.\cite{sanchez84,fontaine94,zunger94,reuter05a} Since the accuracy of the determined lateral interactions should be of the order of $k_BT$ to properly describe the thermal ordering, a concern with this approach has been whether the employed first-principles energetics is actually accurate enough.

Within this context, the present work has a methodological and a materials science motivation. The methodological motivation is to scrutinize both the assumption of exclusively pairwise interactions, and the accuracy with which the first-principles LGH approach can provide the lateral interactions. For this purpose we concentrate on a simple model case, namely the on-surface ordering of atomic adsorbates at a (100) cubic surface, for which extensive studies with model interactions have already been performed.\cite{ree67,binder80,kinzel81,caflisch84,amar84,bak85,bartelt89,liu04} To make contact with a specific material and with experiment, we specifically choose the on-surface adsorption of oxygen at Pd(100), for which detailed experimental data on the ordering behavior is available \cite{chang87}. Since in this system higher oxygen coverages above $\Theta \sim 0.5$\,monolayers [ML, defined with respect to the number of Pd atoms in one layer of Pd(100)] induce structures containing incorporated oxygen \cite{orent82,chang87,chang88,zheng02,todorova03}, we concentrate on the low coverage regime. For this regime, two ordered structures have hitherto been characterized experimentally \cite{chang87,orent82,chang88,zheng02,ertl70,stuve84}: A $p(2 \times 2)$-O structure at 0.25\,ML and a $c(2 \times 2)$-O structure at 0.5\,ML, both with O adsorbed in the on-surface fourfold hollow sites. The material science motivation of our first-principles LGH study is then to extract the lateral interactions operating between the adsorbed O atoms at the surface and to study the ordering behavior they imply. Specifically, this is to see whether we can confirm the experimentally determined ordered structures, as well as the critical temperatures for the order-disorder transition in the low coverage regime.

Presenting a systematic first-principles lattice-gas Hamiltonian expansion, we indeed find the calculated set of lateral interaction energies to be fully consistent with the experimentally reported low coverage phase diagram. Critically discussing the uncertainties of our approach, both with respect to the employed LGH expansion and the underlying DFT energetics, we conclude that they primarily affect the on-site energy. The lateral interaction energies, on the other hand, can be determined with quite high accuracy, which we estimate for the present system to be around 60\,meV. Comparing these interaction energies with those determined previously empirically and using the pairwise interaction approximation demonstrates that the latter assumption introduces an error that is at least as large as this remaining uncertainty when carefully determining the lateral interactions from present-day first-principles calculations.

\section{Theory}

\subsection{Lattice-Gas Hamiltonian}

In order to describe the site-specific adsorption of oxygen atoms on the Pd(100) surface we employ the concept of a lattice-gas Hamiltonian, in which any system state is defined by the occupation of sites in a lattice, and the total free binding energy of any configuration is expanded into a sum of discrete interactions between the lattice sites (see e.g. Refs. \onlinecite{sanchez84,fontaine94,zunger94,reuter05a}). For a one component system with only one site type, this energy reads (with obvious generalizations to multi-site cases)
\begin{eqnarray}
\nonumber
N F_b &=&  F_b^{\rm on-site} \sum_{i} n_{i} \;+\; \sum_{u=1}^{\rm r} V_{u}^{\rm p} \sum_{(i<j)_u} n_{i}n_{j} \;+ \\
  & & \sum_{u=1}^{\rm q}V_{u}^{\rm t} \sum_{(i<j<k)_{u}} n_{i}n_{j}n_{k}
+ \ldots \quad ,
\label{eqLGH}
\end{eqnarray}
where the site occupation numbers $n_{l}$ = 0 or 1 indicate whether site $l$ in the lattice is empty or occupied, with a total of $N$ sites occupied, and $F_b^{\rm on-site}$ is the free energy of an isolated species at the lattice site, including static and vibrational contributions. There are $r$ pair interactions with two-body (or pair) interaction energies $V_{u}^{\rm p}$ between species at $u$th nearest neighbor sites, and $q$ trio interactions with $V^{\rm t}_{u}$ three-body interaction energies. The sum labels $(i<j)_u$ [and $(i<j<k)_u$] indicate that the sums run over all pairs of sites $(ij)$ (and three sites $(ijk)$) that are separated by $u$ lattice constants, and the summation is done such, that each pair (or trio) of occupied sites contributes exactly once to the lattice energy.\cite{footnote}

\begin{figure}[t!]
\scalebox{0.31}{\includegraphics{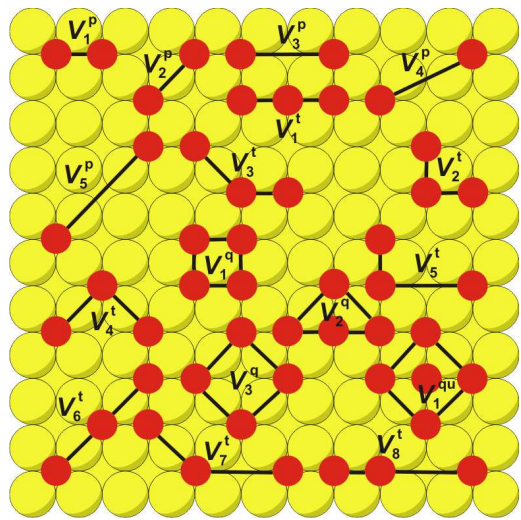}}
\caption{\label{fig1}
(Color online) Top view of the Pd(100) surface, illustrating the considered pool of 17 lateral interactions between O atoms in on-surface hollow sites. $V_{m}^{\rm p}$ ($m=1,2,3,4,5$) are the two-body (or pair) interactions at first, second, third, fourth and fifth nearest neighbor distance. $V^{\rm t}_{n}$, $V^{\rm q}_{n}$, $V^{\rm qu}_{n}$ are considered compact trio, quattro and quinto interactions, respectively. Light grey (yellow) spheres represent Pd atoms, and small dark grey (red) spheres O atoms.}
\end{figure}

Formally, higher and higher order interaction terms (quattro, quinto,...) follow in the infinite expansion of eq. (\ref{eqLGH}). In practice, the expansion must (and can) be truncated after a finite number of terms. Obviously, the judicious choice of which interactions to consider and which ones to neglect must critically affect the reliability of the entire approach. To quantify the impact of this choice on accuracy, we rely on the concept of leave-one-out cross validation (LOO-CV) detailed below to identify the most important interactions out of a larger pool of possible interactions. Figure~\ref{fig1} illustrates the lateral interactions contained in this pool, which range from pair interactions up to the fifth nearest neighbor, via all trio interactions up to second nearest neighbor, to several compact quattro and one quinto interaction. The pool focuses thus on short- to medium-ranged interactions. Interactions at larger distances can be substrate-mediated elastic or of electronic origin \cite{einstein73}, but for the present system we do not expect such interactions to play a role on the accuracy level of interest to this study.

\subsection{Static and vibrational average binding energy}

In order to generate a quantitatively accurate LGH, we parameterize the unknown lateral interaction energies contained in the LGH by first-principles calculations. The central quantities required for this parameterization are computed average free binding energies for a set of ordered configurations of O adsorbed at Pd(100). We write this average free binding energy as
\begin{equation}
F_b(T) \;=\; E_b + F_b^{\rm vib.}(T) \quad ,
\label{bindengdef}
\end{equation}
separating the total and vibrational contributions, $E_b$ and $F_b^{\rm vib.}(T)$ respectively. The former is defined as
\begin{equation}
E_{b} \;=\; - \frac{1}{N_{\rm O}} 
\left[ E^{\rm total}_{\rm O/Pd(100)} - E^{\rm total}_{\rm Pd(100)} - \frac{N_{\rm O}}{2} E^{\rm total}_{\rm O_2 (gas)} \right] \quad .
\label{Obindeng}
\end{equation}
Here $N_{\rm O}$ is the total number of adsorbed O atoms, $E^{\rm total}_{\rm O/Pd(100)}$, $E^{\rm total}_{\rm Pd(100)}$, and $E^{\rm total}_{\rm O_2 (gas)}$ are the total energies of the surface containing oxygen, of the corresponding clean Pd(100) surface, and of an isolated oxygen molecule, respectively.
Since a free O${}_2$ molecule is thus used as the zero reference for $E_b$, a positive binding energy indicates that the dissociative adsorption of O${}_2$ is exothermic at $T=0$\,K.

In order to determine the vibrational contribution to the average free binding we use the phonon density of states $\sigma(\omega)$ and write the vibrational free energy as
\begin{equation}
F^{\rm vib.}(T) \;=\; \int d\omega \; F(T,\omega) \; \sigma(\omega) \quad ,
\end{equation}
where
\begin{eqnarray} \nonumber
F^{\rm vib.}(T,\omega) &=& \hbar \omega \left( \frac{1}{2} + \frac{1}{e^{\beta\hbar\omega} - 1} \right) \\
&&- k_BT \left[ \frac{\beta\hbar\omega}{e^{\beta\hbar\omega} - 1} -
{\rm ln}(1 - e^{-\beta\hbar\omega} \right]
\end{eqnarray}
is the vibrational free energy of an harmonic oscillator of frequency $\omega$.\cite{reuter02c} $k_B$ is the Boltzmann constant, and $\beta = 1/(k_BT)$ the inverse temperature. The vibrational contribution to the average binding energy can then be written in exactly the same way as eq. (\ref{Obindeng}), namely
\begin{eqnarray}
\label{Ofbindeng}
F^{\rm vib.}_{b}(T) &=& - \frac{1}{N_{\rm O}} \left[ 
F^{\rm vib.}_{\rm O/Pd(100)}(T) \; - \right. \\ \nonumber
&& \left. F^{\rm vib.}_{\rm Pd(100)}(T) - \frac{N_{\rm O}}{2} F^{\rm vib.}_{\rm  O_2 (gas)}(T) \right] \,= \\ \nonumber
&& - \frac{1}{N_{\rm O}} \int d\omega \; F^{\rm vib.}(T,\omega) \; 
\left[ \sigma_{\rm O/Pd(100)}(\omega) \; - \right. \\ \nonumber
&& \left. \sigma_{\rm Pd(100)}(\omega) - \frac{N_{\rm O}}{2} \sigma_{\rm  O_2 (gas)}(\omega) \right] \quad .
\end{eqnarray}
To evaluate this contribution in practice one must thus determine the difference of the surface phonon density of states of the adsorbate covered and of the clean surface, $\sigma_{\rm O/Pd(100)}(\omega)$ and $\sigma_{\rm Pd(100)}(\omega)$ respectively, as well as the vibrational frequencies of the gas phase molecule contained in $\sigma_{\rm  O_2 (gas)}(\omega)$.

\subsection{Total Energy Calculations}

The total energies required to evaluate eq. (\ref{Obindeng}) are obtained by DFT calculations within the highly accurate full-potential augmented plane wave + local orbitals (LAPW/APW+lo) scheme \cite{wien2k}, using the generalized gradient approximation (GGA-PBE) \cite{perdew96} for the exchange-correlation functional. All surface structures are modeled in a supercell geometry, employing fully-relaxed symmetric slabs (with O adsorption on both sides of the slab) consisting of five (100) Pd layers with an optimized bulk lattice constant of $a =3.947$\,{\AA} (neglecting bulk zero-point vibrations). A vacuum region of $\ge 10$\,{\AA} ensures the decoupling of consecutive slabs. The LAPW/APW+lo basis set parameters are listed as follows: Muffin tin spheres for Pd and O are $R_{\rm MT}^{\rm Pd} = 2.1$\,bohr and $R_{\rm MT}^{\rm O}= 1.1$\,bohr, respectively, the wave function expansion inside the muffin tin spheres is done up to $l_{\rm max}^{\rm wf}= 12$, and the potential expansion up to $l_{\rm max}^{\rm pot}= 6$. The energy cutoff for the plane wave representation in the interstitial region between the muffin tin spheres is $E_{\rm max}^{\rm wf} = 20$\,Ry for the wave functions and $E_{\rm max}^{\rm pot}=196$\,Ry for the potential. Monkhorst-Pack (MP) grids are used for the Brillouin zone integrations. Specifically, we use a $(12 \times 12 \times 1)$ grid for the calculation of $(1 \times 1)$ surface unit-cells. For the larger surface cells, care is taken to keep the reciprocal space point sampling identical by appropriately reducing the employed {\bf k}-meshes. 

To obtain the total energy of the isolated O$_2$ molecule, we exploit the relation $E^{\rm total}_{\rm O_2(gas)} = 2 E_{\rm O(atom)}^{\rm total} - D$, where $E_{\rm O(atom)}^{\rm total}$ is the total energy of an isolated oxygen atom, and $D$ the theoretical O$_2$ binding energy. The isolated O atom is then calculated spin-polarized, inside a rectangular cell of side lengths $(12 \times 13 \times 14)$\,bohr, $\Gamma$-point sampling of the Brillouin zone and without spherically averaging the electron density in the open valence shell. For $D$ we employ the previously published ultra-converged GGA-PBE value of $6.21$\,eV.\cite{kiejna06}

For the calculations of the adsorbate vibrational modes, the dynamical matrix is set up by displacing the O atom from its equilibrium position in 0.05\,{\AA} steps. Anticipating a good decoupling of the vibrational modes due to the large mass difference between Pd and O, the positions of all atoms in the substrate below the adsorption site are kept fixed in these calculations. The frequencies and normal modes are then obtained by subsequent diagonalization of the dynamic matrix.

\subsection{Monte Carlo simulations}

Once a reliable set of interactions has been established, evaluating the LGH for any configuration on the lattice corresponds merely to performing an algebraic sum over a finite number of terms, cf. eq. (\ref{eqLGH}). Due to this simplicity, the LGH can be employed to evaluate the system partition function. Here this is done by canonical Monte Carlo (MC) simulations for O coverages up to $\Theta=0.5$\,ML. The employed lattice size was $(40 \times 40)$ with periodic boundary conditions. Metropolis sampling used 2000 MC passes per lattice site for equilibration, followed by 10000 MC passes per site for averaging the thermodynamic functions. Increasing any of these numerical parameters led to identical results on the accuracy level of interest to this study, i.e. here primarily critical temperatures that are converged to within 5-10\,K. 

For fixed coverage on the surface, ordered structures are identified by evaluating order parameters sensitive to lateral periodicities. To check on the $(2 \times 2)$ periodicity of the two experimentally characterized ordered structures, we divide the (100) cubic lattice into four interpenetrating sub-lattices $a$, $b$, $c$ and $d$ in a $(2 \times 2)$ arrangement ${a\,b\choose c\,d}$. This allows to separately evaluate in the MC runs the average number of occupied sites in each sub-lattice, $N_a$, $N_b$, $N_c$, and $N_d$, respectively. Using Fourier theory, the order parameter for the $p(2 \times 2)$ structure is then defined as
\begin{equation}
\Psi_{p(2\times 2)} \;=\; \frac{3}{4N_{\rm tot}} \sqrt{N_1^2 + N_2^2 + N_3^2} \quad ,
\label{orderp2x2}
\end{equation}
where $N_1 = N_a + N_b + N_c - N_d$, $N_2 = N_a + N_b - N_c + N_d$, $N_3 = N_a - N_b + N_c + N_d$, and $N_{\rm tot}$ is the total number of sites in the simulation cell. In the same way, the order parameter for the $c(2 \times 2)$ structure is defined as
\begin{equation}
\Psi_{c(2 \times 2)} \;=\; \frac{1}{N_{\rm tot}} \sqrt{ \left( N_a + N_d - N_b - N_c \right)^2 } \quad .
\label{orderc2x2}
\end{equation}
Computing these order parameters as a function of temperature, the critical temperature for the order-disorder transition is defined by the inflection point where $\Psi_{p(2 \times 2)}$ or $\Psi_{c(2 \times 2)}$ go to zero. In parallel, we also derive the critical temperature from evaluating the specific heat, obtaining values that are identical to within 10\,K with those inferred from the order parameters.

\section{First-Principles Lattice-gas Hamiltonian for O at Pd(100)}

\subsection{Energetics of on-surface adsorption}

\begin{figure*}
  \epsfig{bbllx=20,bblly=20,bburx=575,bbury=126,clip=,
          file=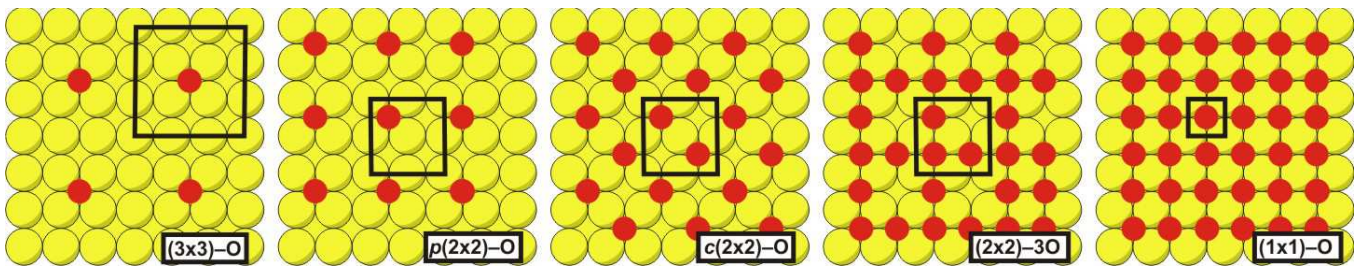,width=1.0\linewidth}
\caption{\label{fig2}
(Color online) Top view of 5 ordered adlayers with O in on-surface hollow sites. The coverage of each configuration from left to right panel is 1/9, 1/4, 1/2, 3/4 and 1 ML, respectively. Light grey (yellow) spheres represent Pd atoms, small dark grey (red) spheres O atoms, and the black lines indicate the surface unit-cells.}
\end{figure*}

\begin{table}
\caption{\label{tab1}
Calculated binding energies $E_b$ (in meV/O atom) for O adsorption in on-surface hollow or bridge sites. The nomenclature and geometric arrangement in the surface unit-cell for the five ordered adlayers are explained in Fig. \ref{fig2}.}
\begin{ruledtabular}
\begin{tabular}{l|ccccc}
         & $(3 \times 3)$-O & $p(2 \times 2)$-O & $c(2 \times 2)$-O & $(2 \times 2)$-3O & $(1 \times 1)$-O \\ 
$\Theta$ & 0.11\,ML         & 0.25\,ML          & 0.50\,ML         & 0.75\,ML & 1.00\,ML \\ \hline
hollow   & 1249 & 1348 & 1069 & 643 & 344 \\
bridge   & 1024 &  961 &  801 & 573 & 378 \\
\end{tabular}
\end{ruledtabular}
\end{table}

Owing to the tendency of oxygen atoms to prefer highly coordinated binding sites at late transition metal surfaces, the high-symmetry fourfold hollow sites appear as the most likely adsorption sites at Pd(100). On the other hand, one cannot exclude {\em a priori} that the twofold bridge sites are not also metastable, i.e. local minima of the potential energy surface. To test this, we slightly displaced a bridge site O adatom in a $p(2 \times 1)$ configuration laterally towards a neighboring hollow site. The resulting forces relaxed the adatom back to the ideal bridge position, so that at least in this configuration the bridge site is not just a mere transition state, i.e. a saddle point of the potential energy surface. As this might also be true for bridge site adsorption in other (local) O adatom arrangements, we calculated the binding energetics of O atoms in the fourfold hollow and in the twofold bridge sites for 5 different ordered overlayers spanning the coverage range up to one ML. The periodicities of these overlayers are explained for the case of hollow site adsorption in Fig. \ref{fig2}, and Table \ref{tab1} summarizes the calculated binding energies.

For lower coverages the fourfold hollow site is energetically clearly more stable, which suggests an insignificant contribution of bridge sites to the ordering behavior at coverages up to around 0.5\,ML, even if the latter are always metastable sites. Although the reversal of the energetic order between hollow and bridge sites at $\Theta =1$\,ML seen in Table \ref{tab1} is intriguing, it clearly occurs in a coverage range where surface oxide formation and eventually three-dimensional oxide cluster growth takes place.\cite{zheng02,todorova03} Since our interest lies in the on-surface ordering behavior at low coverages, we will therefore focus the LGH expansion for the moment exclusively on adsorption into the fourfold hollow sites, and return to the role of bridge site O atoms in Section IIID. 

\begin{figure}
\scalebox{0.36}{\includegraphics{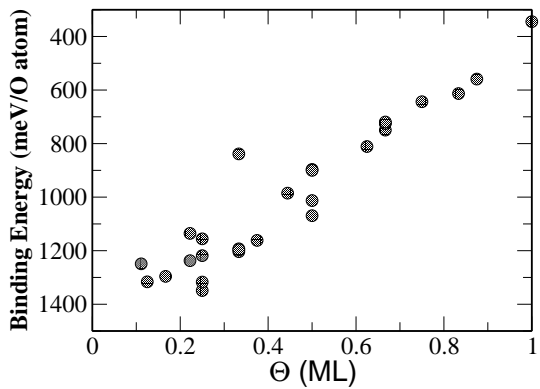}}
\caption{\label{fig3}
(Color online) Coverage ($\Theta$) dependence of the calculated DFT binding energies of 25 ordered configurations with O atoms in on-surface hollow sites.}
\end{figure}

As the next step in the LGH parameterization, average binding energies for different ordered configurations with O atoms in on-surface hollow sites and with surface unit-cells up to $(3 \times 3)$ were correspondingly computed. Despite our focus on the lower coverage regime, the set does comprise structures with coverages up to 1\,ML, since these structures are required during the LGH parameterization to determine in particular the higher-order many-body interactions occurring in (locally) denser adatom arrangements. In most cases configurations that we initially prepared with high coverage of on-surface O atoms were truly metastable, in the sense that they relaxed into geometries where all O atoms remained in the on-surface hollow sites. However, some structures directly relaxed into geometries with O incorporated below the first Pd layer. Since the O-O interaction in these structures does not correspond to the physical situation we want to describe (and would thus mess up the chosen LGH), we excluded these configurations from our set (using a strongly enlarged Pd first interlayer distance as criterion). This left a set of 25 configurations with on-surface O atoms, which was subsequently used in the parameterization of the LGH. The binding energy data of this set is compiled in Fig. \ref{fig3}, and indicates already overall strongly repulsive lateral interactions, which reduce the binding energy with increasing coverage by up to 1\,eV.

\subsection{Lateral interactions}

In order to determine the lateral interaction energies in the LGH, we employ eq. (\ref{eqLGH}) to write down the LGH expression for each of the ordered configurations calculated by DFT (including the interactions with the periodic images in the neighboring cells). Neglecting the vibrational contributions in eq. (\ref{bindengdef}) for the moment, we equate the right hand side of eq. (\ref{eqLGH}) with $N_{\rm O} E_b$ for the corresponding configurations and arrive at a system of linear equations that can be solved for the unknown values of the lateral interaction energies. The crucial aspects of this procedure are therefore i) the number and type of interactions included in the LGH expansion, and ii) the number and type of ordered configurations that are computed with DFT to determine the values of these interaction energies. In the following we show how i) is addressed by leave-one-out cross-validation, and ii) is aided by a search for the LGH ``ground state'' structures and an iterative refinement of the input structure set.

\subsubsection{Leave-one-out cross-validation (LOO-CV)}

In a truncated LGH expansion with finite ranged interactions, sparse configurations will exhibit a lattice energy that is simply $N_{\rm O}$ times the on-site energy, as soon as all adsorbed species have distances from each other that exceed the longest range considered interaction. We therefore fix the on-site energy $E_b^{\rm on-site}$ in eq. (\ref{eqLGH}) to be the DFT binding energy computed for 1/9\,ML coverage, cf. Fig. \ref{fig2} and Table \ref{tab1}. In this particular $(3 \times 3)$-O configuration, the minimum distance between O adatoms is 8.37\,{\AA}, i.e. six nearest neighbor sites away. This is larger than the farthest reaching interaction contained in our pool of lateral interactions, cf. Fig. \ref{fig1}, so that fixing $E_b^{\rm on-site}$ to the 1/9\,ML $(3 \times 3)$-O binding energy should prevent fitting noise into this parameter. 

To get some guidance as to which could be the leading lateral interactions to be included in the LGH expansion, we estimate the predictive power of the LGH by the concept of leave-one-out cross-validation \cite{zunger94,shao93,zhang93,walle02,borg05}. For a given set of LGH interactions the cross-validation (CV) score is calculated as
\begin{equation}{\label {ecv}}
{\rm CV} \;=\; \sqrt{ \frac{1}{M} \sum_{i=1}^{M}
\left( E_b^{\rm DFT}(i) - E_b^{{\rm LGH^\prime}}(i) \right)^2 } \quad .
\end{equation}
Here, the sum runs over the remaining $M=24$ ordered configurations that were calculated with DFT (apart from the 1/9\,ML coverage structure used already for $E_b^{\rm on-site}$), and which have DFT calculated binding energies $E_b^{\rm DFT}(i)$. The quantity $E_b^{{\rm LGH^\prime}}(i)$ of the $i$th configuration, on the other hand, is evaluated from the LGH expression for this configuration, cf. eq. (\ref{eqLGH}), where the values of the considered lateral interactions are obtained from least-squares fitting to the DFT energies of the remaining $M-1$ calculated configurations, i.e. leaving exactly the $i$th configuration out of the fit. This way, the CV score is intended to be a measure of the predictive power of a LGH expansion considering a given set of lateral interactions. In general, one would expect sets containing too few interactions to be too inflexible and thus leading to a high CV score, whereas sets containing too many interactions as loosing their predictive power through overfitting and thereby also leading to a high CV score. The hope is thus to identify the optimum set of considered interactions as that set that minimizes the CV score. 

\begin{table*}
\begin{center}
\caption{\label{tab2}
List of the sets containing $m$ lateral interactions, together with their CV scores and the values determined for the lateral interaction energies (no entry at a position in the table means that this interaction is not contained in the set. Lateral interactions shown in Fig. \ref{fig1}, but not shown here are never selected out of the pool). Negative values for the interaction energies indicate repulsion, positive values attraction. The sets shown are those that minimize the CV score among all possible sets containing $m$ lateral interactions out of the pool of 17 shown in Fig. \ref{fig1}. The first line for each set corresponds to the data obtained by fitting to 24 ordered DFT configurations, while the second line is obtained after adding 2 additional DFT configurations to the fit (see text). Units for the CV score and lateral interactions are meV.}
\vspace{0.1 cm}
\begin{tabular}{c|c||c|c|c|c|c||c|c|c|c|c|c|c||c|c}
m & CV & \multicolumn{5}{c||}{pair} & \multicolumn{7}{c||}{trio} & \multicolumn{2}{c}{quattro} \\
  & score & $V^{\rm p}_{1}$ & $V^{\rm p}_{2}$ & $V^{\rm p}_{3}$ & $V^{\rm p}_{4}$ & 
$V^{\rm p}_{5}$ & $V^{\rm t}_{1}$ & $V^{\rm t}_{2}$ & $V^{\rm t}_{3}$ & 
$V^{\rm t}_{4}$ & $V^{\rm t}_{5}$ & $V^{\rm t}_{6}$ & $V^{\rm t}_{7}$ & $V^{\rm q}_{1}$ & $V^{\rm q}_{2}$\\ \hline
7  & 31 & $-$344 & $-$130& 52 &14 & &$-$120&132 & $-$54 & & &       & & &     \\ 
   & 32 & $-$338 & $-$126& 50 &16 & &$-$114&117 & $-$51 & & &       & & &     \\ \hline
8  & 20 & $-$324 & $-$126& 50 &16 & &$-$138&120 & $-$57 & & &       & & & 102  \\ 
   & 22 & $-$314 & $-$122& 46 &18 & &$-$138&108 & $-$57 & & &       & & & 120  \\ \hline
9  & 16 & $-$292 & $-$90 & 50 &10 & &$-$168& 60 & $-$48 & & & $-$51 & & & 120  \\ 
   & 16 & $-$296 & $-$92 & 50 &10 & &$-$162& 63 & $-$48 & & & $-$51 & & & 114  \\ \hline
10 & 17 & $-$290 & $-$92 & 50 &10 & &$-$168& 60 & $-$48 & & 1       & $-$51 & &   & 120 \\  
   & 17 & $-$298 & $-$92 & 50 &10 & &$-$162& 69 & $-$51 & &          & $-$54 & & $-$42 & 120 \\ \hline
11 & 18 & $-$284 & $-$90 & 46 &12 & 6 & $-$171& 57 & $-$51 & & $-$3  & $-$54 & & & 132 \\
   & 18 & $-$296 & $-$92 & 48 &10 & 2 & $-$165& 69 & $-$48 & &       & $-$54 & & $-$42 & 126 \\ \hline
12 & 20 & $-$292 & $-$70 & 50 & 8 & 2 & $-$168& 486& $-$264& 414 &  & $-$36 & $-$216 & & 132 \\
   & 19 & $-$296 & $-$92 & 48 &10 & 2 & $-$165& 69 & $-$51 &     & $-3$ & $-$54 &     & $-$48 & 126 \\
\end{tabular}
\end{center}
\end{table*}

Within this approach, we evaluate the CV score for any set of interactions out of the larger pool of 17 lateral interactions shown in Fig. \ref{fig1}. Table \ref{tab2} summarizes these scores subdivided into the optimum sets containing $m$ lateral interactions, i.e. listed are the sets that yield the lowest CV score for any arbitrary combination of $m$ lateral interactions out of the total pool of 17. For these sets, we then determine the values of the considered lateral interactions by least-squares fitting to the computed DFT binding energies of all $M=24$ ordered configurations and also include them in Table \ref{tab2}. The minimum CV score reached indeed decreases initially upon adding more interactions to the set, and then increases again for sets containing more than 10 interactions. Another gratifying feature is that almost always the same interactions are picked out of the pool, i.e. the optimum set for $m+1$ interactions corresponds mostly to those interactions already contained in the optimum set for $m$ interactions plus one additional one. Only very rarely is an interaction that is contained in the optimum $m$ set not selected in the optimum $m+1$ set. And if this happens, this concerns lateral interactions for which only very small values are determined, and which are thus anyway not meaningful within the uncertainties of our approach. Also in a physical picture, the determined values for the lateral interactions appear quite plausible for $m$ up to 11. The pairwise interactions decrease with increasing distance, and the leading higher-order trio and quattro interactions are smaller in size than the most dominant nearest-neighbor pair interaction. The quinto interaction contained in the pool of possible interactions is never selected. In contrast, the $m =12$ set shows already clear signs of linear dependencies, with some trio interactions suddenly exhibiting very large values. This continues for expansions containing even more interactions (not listed in Table \ref{tab1}), which exhibit more and more obviously meaningless lateral interaction energies.

Another equally important feature of the expansions up to $m = 11$ is the stability of the determined lateral interaction values against adding further interactions to the set. In particular for the optimum sets close to the overall CV minimum, i.e. for $m$ equal to 9 or 10, adding another lateral interaction to the set changes the values for the dominant interactions by less than 2\,meV. A similar behavior is obtained for another test to which we subject our expansion: Having calculated another two DFT configurations, we increased the set of DFT configurations employed in the fit to $M=26$ and repeated the entire CV score evaluation. The results are also included in Table \ref{tab2} and show only minimal variations for the sets up to $m=11$. Almost always the same lateral interactions are picked out of the pool and also their values change by less than 10\,meV compared to the previous procedure employing 24 DFT configurations. 
For the set $m=12$, adding the two DFT input energies removes the linear dependencies and brings the set back in line with the other sets with fewer interactions. These findings suggest that the expansions are also stable against adding further DFT configurations and we finally identify the set containing 9 lateral interactions and using 24 DFT configurations to determine their values as our optimum LGH expansion.

\subsubsection{Ground state search}

Before moving to the ordering behavior at finite temperatures, a crucial test for the validity of the LGH expansion is that it gives the correct ordered ground states at $T = 0$\,K, i.e. the lowest-energy structures at a given coverage. Here, this refers in particular to the ground states predicted by the DFT energetics, since the latter is the input with which the LGH expansion must be consistent. Obviously, if the energetic order of competing configurations is wrong at the DFT level (e.g. due to the employed approximate xc-functional), there is no hope that a correct LGH expansion could cure this problem. To this end, it is useful to replot the DFT database shown in Fig. \ref{fig3} in form of a formation energy plot \cite{zunger94}. Formation energies $\Delta E_f$ are in general defined as an excess energy with respect to the equivalent amounts of pure constituents. For the present case of on-surface O adsorption in the coverage range below 1\,ML we therefore define
\begin{eqnarray}
\nonumber
\Delta E_f &=& \frac{1}{N_{\rm tot}} \left[ E^{\rm total}_{\rm O/Pd(100)} - (1 - \Theta) E^{\rm total}_{\rm Pd(100)} - \right. \\
& & \left. \quad \quad \Theta E^{\rm total}_{\rm (1 \times 1)-O/Pd(100)} \right] \quad .
\end{eqnarray}
As in eq. (\ref{Obindeng}), $E^{\rm total}_{\rm O/Pd(100)}$ is the total energy for a specific adsorbate configuration with $N_{\rm O}$ O atoms per surface unit-cell (corresponding to a coverage $\Theta= N_{\rm O}/N_{\rm tot}$ with $N_{\rm tot}$ the number of sites per surface unit-cell), $E^{\rm total}_{\rm Pd(100)}$ is the total energy of the clean surface, and $E^{\rm total}_{\rm (1 \times 1)-O/Pd(100)}$ is the total energy of the full monolayer $(1 \times 1)$-O configuration. With this definition, $\Delta E_f$ reflects the relative stability of a particular configuration with respect to phase separation into a fraction $\Theta$ of the full monolayer configuration and a fraction $(1 - \Theta)$ of clean surface, and we can relate it to the binding energy of the configuration by
\begin{equation}
\Delta E_f \;=\; \Theta \left[ E_{b,{\rm O/Pd(100)}} - E_{b,{(1 \times 1)-{\rm O/Pd(100)}}} \right]
\quad .
\label{deltaf_2}
\end{equation}

\begin{figure}
  \epsfig{bbllx=20,bblly=20,bburx=575,bbury=411,clip=,
          file=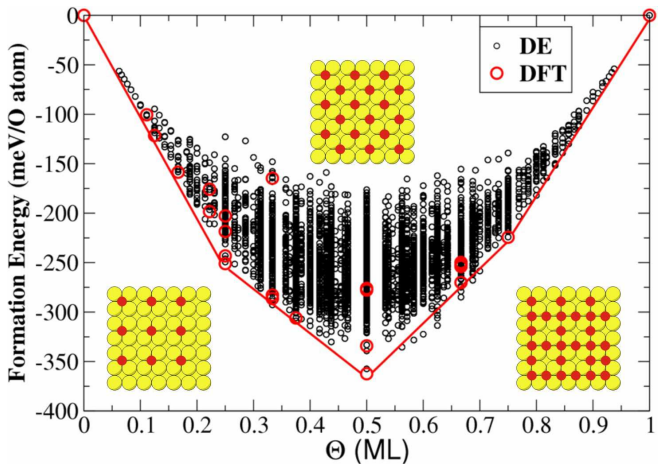,width=1.0\columnwidth}
\caption{\label{fig4}
(Color online) Formation energies $\Delta E_f$ as computed with DFT for the 25 ordered configurations shown in Fig. \ref{fig2} (big (red) circles), as well as for all configurations obtained by direct enumeration and using the LGH expansion (small (black) circles), see text. The red line is the convex hull for the DFT data, identifying three ordered ground states shown as insets: $p(2 \times 2)$-O at $\Theta = 0.25$\,ML (left inset), $c(2 \times 2)$-O at $\Theta = 0.5$\,ML (top inset), and $(2 \times 2)$-3O at $\Theta = 0.75$\,ML (right inset). Pd = large, light (yellow) spheres, O = small, dark (red) spheres.}
\end{figure}

Plotting $\Delta E_f$ versus coverage as done in Fig. \ref{fig4} allows to identify the ground states, i.e. lowest-energy ordered phases, that are predicted by the present DFT data set, as those lying on the convex hull (or so-called ground state line).\cite{zunger94} Any structure that yields a formation energy that is higher than this ground state line is unstable against decomposition into the two ordered configurations represented by the two closest lying corner points on the convex hull. As apparent from Fig. \ref{fig4} the convex hull formed by the DFT data set exhibits three ordered ground states (apart from the trivial ones at the ends of the considered coverage range). Consistent with existing experimental data, these are the $p(2 \times 2)$-O ordered phase at $\Theta = 0.25$\,ML, and the $c(2 \times 2)$-O ordered phase at $\Theta = 0.5$\,ML. A third ordered structure, $(2 \times 2$)-3O at $\Theta = 0.75$\,ML, is at best metastable, since it falls already in the coverage range above $\sim 0.5$\,ML, for which surface oxide formation sets in.

Using eq. (\ref{deltaf_2}) we can also evaluate formation energies using the binding energies obtained from the LGH expansion. Since the evaluation of the latter is numerically significantly less demanding, we can sample a much larger configuration space in this case. To this end, we directly enumerate all combinatorially possible ordered structures in surface unit-cells of any symmetry and with a surface area smaller or equal to a $(4 \times 4)$ cell and with O coverages up to 1\,ML. The corresponding data points are also shown in Fig. \ref{fig4}. If we first focus on the coverage range up to 0.5\,ML, we find the obtained LGH data to be fully consistent with the DFT ground state line. Namely, there is no structure predicted by the LGH expansion that would have an energy that is lower than the DFT convex hull, and the LGH Hamiltonian yields therefore exactly the same ordered ground states as the DFT input data. 

In fact, this is not a coincidental result demonstrating the reliability of the achieved LGH expansion, but the end product of an iterative procedure. We used the consistency with the DFT ground state line as another criterion to judge whether more DFT ordered configurations are required as input to the LGH expansion. Initially we had started with a smaller number of DFT configurations as the set discussed above. Having gone through the same CV score evaluation, we had identified an optimum LGH expansion, but had then obtained LGH data points in the direct enumeration leaking below the DFT ground state line. Interpreting the corresponding structures as important input to the LGH expansion, we would ideally calculate them with DFT and add them to the DFT data set used for the LGH expansion. This was, unfortunately, not always possible, when the structures predicted by the LGH had surface unit-cells that exceeded our computational capabilities. In such cases, we looked for other structures in smaller unit-cells, which still contained what we believed were the relevant motifs and computed those with DFT. This procedure was repeated several times, each time adding new structures to the DFT data base, until the present consistent result was obtained.

In the coverage range above $\sim 0.5$\,ML, the situation is not that perfect. As apparent from Fig. \ref{fig4} there are still several LGH structures slightly below the DFT ground state line. Unfortunately, further improvement along the sketched lines is inhibited by the above described propensity of structures in this coverage range to directly relax into geometries with O incorporated below the first Pd layer. This renders it very tough to provide new on-surface O/Pd(100) structures to the data base and improve on the present LGH expansion. Although not completely satisfying, we therefore contend ourselves with the achieved expansion. Particular care should therefore be exerted, when aiming to use it to describe the higher coverage regime, since denser adatom arrangements can presumably not be fully reliably described. However, due to the overall strongly repulsive interactions, the local occurrence of such denser arrangements at lower coverages is rather unlikely in the MC simulations. Correspondingly, we do expect the results obtained from our expansion to be reliable for the coverage range below $\sim 0.5$\,ML, on which we focus in the present work.

\subsection{Order-disorder transition}

Having established the ground state ordered structures we proceed to study the ordering behavior at finite temperatures. Experimentally, this was investigated in the coverage range up to 0.6\,ML by Chang and Thiel \cite{chang87}. For defined initial coverages at the surface, they identified the presence of ordered phases at the surface by monitoring LEED superstructure spots corresponding to the different periodicities, and the critical temperatures $T_c(\Theta)$ for the order-disorder transition were determined by the inflection point of the vanishing spot intensities at increasing temperatures. Avoiding the O-induced reconstruction at higher coverages, we focus here on the data in the coverage range $\Theta < 0.35$\,ML, in which the $p(2 \times 2)$ or a coexistence of $p(2 \times 2)$ and $c(2 \times 2)$ phases form the ordered structures at low temperatures, cf. Fig. \ref{fig4}. For this coverage range, Chang and Thiel determined the onset of desorption in their ultra-high vacuum experiments at much higher temperatures than the order-disorder transition. \cite{chang87} From this we assume that in the experiments, the coverage at the surface remained essentially constant at the initially prepared coverage value for all temperatures up to the critical temperatures for the order-disorder transition. 

\begin{figure}
\scalebox{0.40}{\includegraphics{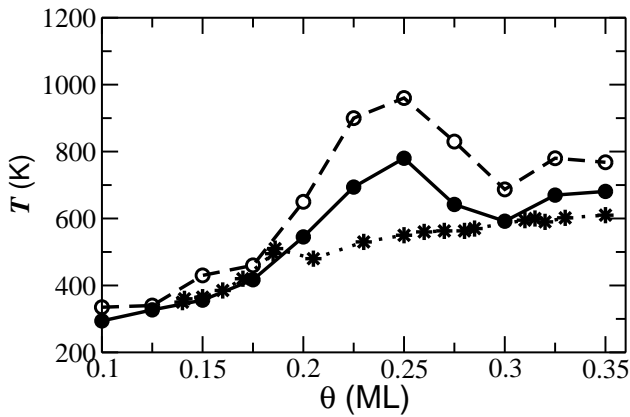}}
\caption{\label{fig5}
$\Theta-T$ diagram of the critical temperatures $T_c$ for the order-disorder transition. Shown with stars are the experimental data from Ref. \onlinecite{chang87}, and with solid circles the values obtained when using the optimum LGH expansion. Additionally shown by empty circles are the values obtained when using the LDA as exchange-correlation functional in the DFT calculations (see text).}
\end{figure}

The experimental conditions are simulated by canonical MC runs for fixed coverages and at various temperatures. With the definitions in eqs. (\ref{orderp2x2}) and (\ref{orderc2x2}), our order parameters are equivalent to LEED spot intensities, so that the determined critical temperatures for the order-disorder transition can be directly compared to the experimental values. Figure \ref{fig5} shows the $T_c(\Theta)$ curve obtained with the optimum LGH expansion together with a reproduction of the experimental data. Overall, we observe very good agreement, both with respect to the absolute temperature values and the trend of increasing critical temperatures with coverage. The largest deviation of about 250\,K results at $\Theta = 0.25$\,ML, where theory predicts a small peak in the critical temperature, which is absent in the experimental data and which we discuss in Section IVB below. Apart from this feature, the agreement with the experimental $T_c(\Theta)$ data is quite satisfying, if not quantitative.

\subsection{Population of bridge sites}

The LGH+MC simulations up to now have exclusively focused on O adsorption into the fourfold on-surface hollow sites. The already good agreement obtained with existing experimental data, together with the significantly lower stability of the energetically next favored high-symmetry bridge sites apparent in Table \ref{tab1}, seem to suggest that the on-surface low coverage ordering can indeed be understood in terms of the most stable hollow sites only. To verify the implied negligible population of (possibly metastable) bridge sites, even up to the critical temperatures of the order-disorder transition, we proceed by including these sites into the LGH expansion. Since the intention is at this point only to check on the influence of a population of these sites, we consider a reduced pool of possible lateral interactions between O atoms adsorbed in bridge sites, consisting of the equivalents to the hollow-hollow pair and trio interactions shown in Fig. \ref{fig1}. Due to the twofold symmetry of the bridge sites, two different forms at the same interatomic distance exist for some of the interactions, and in addition there is a lateral interaction $V_0^{\rm p}$ at the very short distance of $a/2$ between O atoms sitting in immediately adjacent bridge sites coordinated to the same Pd atom. 

However, when computing with DFT configurations containing such closely neighboring O atoms at a site distance of $a/2$, we always found them to be unstable against relaxation. During the geometry optimization the O atoms moved to sites further apart, indicating a sizable repulsive $V_0^{\rm p}$. Focusing therefore on 18 ordered configurations that do not contain O adatoms at such close distance, the best sets with a varying total number of lateral interactions are determined via LOO-CV in the same manner as for the hollow-hollow interactions. Similar to the results in Table \ref{tab2}, the different expansions yield consistently the same two dominant lateral interactions, namely the first-nearest neighbor pair $V_1^{\rm p}({\rm bridge-bridge})$ and second-nearest neighbor pair $V_2^{\rm p}({\rm bridge-bridge})$ interactions. Both are largely repulsive, with $V_1^{\rm p}({\rm bridge-bridge}) \approx -400$\,meV and $V_2^{\rm p}({\rm bridge-bridge}) \approx -120$\,meV. Turning to the leading lateral interactions between O atoms adsorbed in hollow and bridge sites, we again found that structures with O atoms in directly adjacent bridge and hollow sites at the very short distance of $a/\sqrt{8}$ are not stable against structural relaxation. As for the other interactions, this time the first-nearest neighbor pair interaction and one compact trio interaction turn out to be dominant in the LOO-CV based LGH expansion procedure. They are also largely repulsive, with values of $\approx -240$\,meV and $-280$\,meV, respectively.

We therefore approximately consider the interactions involving bridge site O atoms in the MC simulations by excluding configurations with O atoms at the above described shortest $a/2$ and $a/\sqrt{8}$ distances between bridge-bridge and bridge-hollow site pairs, respectively. In addition, the two next dominant lateral interactions for bridge-bridge and bridge-hollow pairs are explicitly accounted for using the values stated above. The consecutive MC simulations show virtually no change in the ground state ordered structures and critical temperatures in the coverage range up to $\Theta = 0.35$\,ML. The overall largely repulsive interactions together with the significantly less stable on-site energy compared to adsorption in the hollow sites, efficiently prevents any significant population of bridge sites. For all temperatures up to the order-disorder transition, we find less than 10\,\% of the available bridge sites occupied with O atoms, with the highest populations obtained for the larger coverages in the considered range. To make sure that these results are not affected by the uncertainty in the approximately determined lateral interactions, we varied the value of each of the four lateral interactions by $\pm 100$\,meV and each time reran the MC simulations. This had no influence on the findings, so that we do not expect them to be invalidated by the crude way of how the bridge-bridge and bridge-hollow interactions are considered. Instead, we conclude that up to coverages of $\Theta \approx 0.35$\,ML and up to the critical temperatures for the order-disorder transition, a population of (possibly metastable) bridge sites plays no role for the on-surface ordering behavior.

\section{Accuracy of first-principles lateral interactions}

The agreement with the experimental low coverage phase diagram (ground state structures and critical temperatures) suggests that the determined {\em set} of first-principles lateral interactions is quite reliable. In order to get a better understanding of the explicit uncertainties of the different parameters in this set, we return to a critical discussion of all approximations entering the LGH approach, and scrutinize their influence on the lateral interaction values. Uncertainties arise on the one hand side due to the truncated LGH expansion and the finite number of configurations employed to parameterize it, and on the other hand due to the approximate first-principles energetics, both with respect to total and vibrational free energy contributions. 

\subsection{Uncertainties in the LGH expansion procedure}

Table \ref{tab2} provides detailed information about the influence of most approximations in the LGH expansion procedure. Inspecting the basically indistinguishable CV score for the expansions with $m =$\,9, 10, and 11, one might take the scatter in the correspondingly extracted lateral interactions as a rough measure for the uncertainty introduced by truncating the LGH expansion after a finite number of terms. Concerning the finite number of DFT configurations employed in the parameterization, the achieved consistency of the DFT and LGH ground state line illustrated in Fig. \ref{fig4} gives some indication as to the minimum number of configurations that is required. Correspondingly, the differences in the lateral interaction values determined when extending this minimum set by two further configurations (upper vs. lower line for each expansion in Table \ref{tab2}) may be taken as reflecting the uncertainty due to employing a finite number of DFT configurations in the parameterization.

This leaves as remaining {\em ad hoc} feature of our expansion procedure the decision to not include the on-site energy into the fit, but instead fix it at the value of the sparsest DFT configuration computed, i.e. the $(3 \times 3)$-O structure at 1/9\,ML. To this end, we redid all LGH expansions in Table \ref{tab2} using the same LOO-CV procedure described above, but this time including the $(3 \times 3)$-O structure into the set of DFT configurations and including the on-site energy $E_b^{\rm on-site}$ into the fit. The results are remarkably consistent, in the sense that the obtained lateral interactions differ in all cases by less than 15\,meV from the values compiled in Table \ref{tab2}, and for the expansions with $m=$ 9, 10, 11 the on-site energy is in fact determined at values that are within 15\,meV of the computed binding energy of the $(3 \times 3)$-O structure. For expansions using fewer lateral interaction figures ($m < 9$) this becomes progressively worse, and the increasing inflexibility of the few-interaction expansions starts to shift errors between the on-site energy and the lateral interactions in an uncontrolled way. We therefore conclude that for expansions with a sufficient number of interaction figures it apparently makes very little difference to include or not include the on-site energy into the fit; the expansions are stable in this respect. In view of the significantly different inaccuracies in the determination of the on-site energy and lateral interactions discussed below, we nevertheless prefer to fix the on-site energy at the value of the sparsest DFT configuration computed. 

\begin{figure}
\scalebox{0.40}{\includegraphics{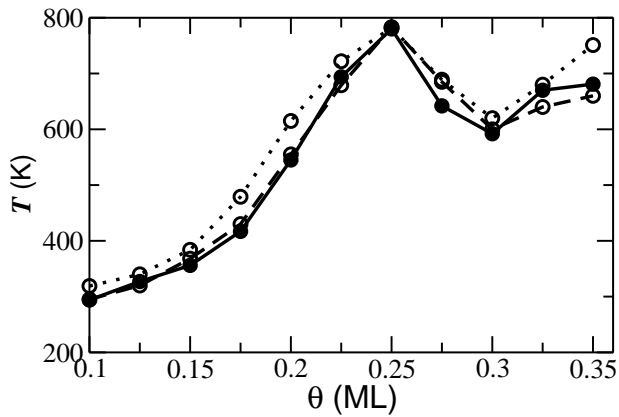}}
\caption{\label{fig6}
$\Theta-T$ diagram of the critical temperatures $T_c$ for the order-disorder transition. Compared are the curves obtained when using the first-principles lateral interactions for different LGH expansions compiled in Table \ref{tab2} and each time using 24 DFT configurations in the parameterization: Solid line, full circles: $m = 9$; dashed line, empty circles: $m = 10$; dotted line, empty circles: $m = 11$.}
\end{figure}

Summarizing the discussion on the uncertainties in the LGH expansion and parameterization procedure, we estimate the uncertainty introduced by the various approximations to be of the order of 10-20\,meV in the dominant lateral interactions. When using the first-principles lateral interactions to determine quantities describing the mesoscopic ordering behavior, this uncertainty needs to be taken into account. Figure \ref{fig6} illustrates this for the determined critical temperatures for the order-disorder transition by comparing the results obtained for the different LGH expansions with $m=$\,9, 10, and 11 in Table \ref{tab2}, for which the extracted lateral interactions exhibit a scatter of about the estimated magnitude. The critical temperatures show a variation of less than 100\,K and the overall form of the $T_c(\Theta)$-curve is almost untouched. The systematics of the present LGH expansion procedure provides thus an error-controlled link between the electronic structure calculations and the mesoscopic statistical simulations, which allows to assess which quantities can be determined with which uncertainty for a given accuracy of the underlying first-principles energetics.

\subsection{Uncertainties in the first-principles energetics}

The energetic parameters in the LGH expansion in eq. (\ref{eqLGH}) comprise total and vibrational free energy contributions. Uncertainties in our approach arise therefore out of the treatment of the vibrational free energy contribution and the approximate DFT energetics, where the latter contains uncertainties due to the numerical setup and due to the approximate exchange-correlation functional. We will discuss these three sources of uncertainties subsequently.

\subsubsection{Vibrational free energy contribution}

Separating each of the energetic parameters in the LGH (i.e $F_b^{\rm on-site}$, $V^{\rm p}_m$, $V^{\rm t}_m$, ...) into a total energy term and a term due to the vibrational free energy contribution, one arrives at a LGH expansion for the total binding energies and a LGH expansion for the vibrational free binding energies. In Section IIIB the vibrational part was completely neglected by equating the right hand side of eq. (\ref{eqLGH}) solely to the $N_{\rm O} E_b$ of the different computed configurations. The formally correct procedure would be to evaluate for each configuration also the vibrational free energy contribution to the binding energy, $F^{\rm vib.}_b(T)$, and consider it in the parameterization. As apparent from eq. (\ref{Ofbindeng}), what determines this vibrational contribution are the changes of the vibrational modes during adsorption. Since this will predominantly concern the adsorbate-derived vibrational modes, we estimate the order of magnitude of this contribution by the zero point energy ($E^{\rm ZPE}$) difference arising from the change of the O$_2$ stretch frequency, $\omega_{\rm O_2(gas)}$, to the O-substrate stretch mode, i.e. we approximate eq. (\ref{Ofbindeng}) roughly with
\begin{eqnarray}
F^{\rm vib.}_{b}(T) &\sim& - \frac{1}{N_{\rm O}} \left[ 
E^{\rm ZPE}_{\rm O/Pd(100)} \; - \frac{N_{\rm O}}{2} E^{\rm ZPE}_{\rm  O_2 (gas)} \right] \\ \nonumber
&\sim& - \frac{1}{N_{\rm O}} \sum_{i=1}^{N_{\rm O}} \frac{\hbar}{2}
\left[ \omega_{\rm O/Pd(100)}(i) \; - \frac{1}{2} \omega_{\rm  O_2(gas)} \right] \quad ,
\label{OZPEdiff}
\end{eqnarray}
where $\omega_{\rm O/Pd(100)}(i)$ is the O-substrate stretch frequency of the $i$th adsorbed O atom in the corresponding configuration. Within the frozen substrate approximation we calculate the latter stretch mode in good agreement to the experimental data \cite{simmons91} as $\hbar \omega_{p(2 \times 2)} = 50$\,meV in the $p(2 \times 2)$-O configuration and $\hbar \omega_{c(2 \times 2)} = 33$\,meV in the $c(2 \times 2)$-O configuration. Compared to the computed stretch frequency of $\hbar \omega_{\rm O_2(gas)} = 190$\,meV of the O$_2$ molecule, this yields an estimated vibrational contribution to the binding energy of $-(\hbar \omega_{p(2 \times 2)} - \hbar \omega_{\rm O_2(gas)}/2)/2 = 23$\,meV and $-(\hbar \omega_{c(2 \times 2)} - \hbar \omega_{\rm O_2(gas)}/2)/2 = 31$\,meV per O atom for the two configurations, respectively. In the LGH expansion these contributions get separated into a non-coverage dependent part, which enters the on-site energy, and a coverage-dependent part, which enters the lateral interactions. Taking the difference between the estimated $p(2 \times 2)$-O and the $c(2 \times 2)$-O vibrational free energy contributions as a measure of the coverage dependence we thus arrive at an uncertainty of $\sim 5$\,meV in the lateral interactions and an uncertainty of $\sim 30$\,meV in the on-site energy due to the neglect of vibrational contributions in our approach. The on-site energy is thus much more affected by the predominantly non-coverage dependent shift in vibrational frequency from the O$_2$ stretch to the O-substrate stretch mode.

\subsubsection{Numerical uncertainties in the DFT energetics}

Turning to the effect of the approximate DFT total energies, we first distinguish between numerical inaccuracies which arise out of the finite basis set, the k-point sampling or the chosen supercell geometries, and the more fundamental uncertainty due to the employed approximate xc functional. In principle, the numerical inaccuracies can be reduced to whatever limit desired, although this may quickly lead to unfeasible computations in practice. In this respect, the finite slab and vacuum thicknesses employed in our supercell setup are the end result of extensive test calculations, in which these quantities were progressively increased until the relative binding energies were converged to the desired limit of $\pm 10$\,meV/O atom. An important point to note concerning the k-point sampling is that compatible k-meshes must be employed in calculations involving different surface unit-cells. Compatible in this respect means that always the same points within the Brillouin zone are sampled, even when the latter changes its size due to the different periodicity of the real-space surface unit-cell. As long as this is considered, we found only a negligible influence on the LGH parameters, when increasing the k-point density to higher values than the one of our standard computational setup described in Section IIC. This leaves as most notable numerical approximation the finite energy cutoff for the plane wave representation in the interstitial region between the muffin tin spheres in the LAPW/APW+lo scheme. To this end, we repeated the LGH expansion with the optimum $m=9$ lateral interaction figure set, cf. Table \ref{tab2}, using as input the 24 DFT configurations with total energies computed self-consistently at cutoffs of 18\,Ry, 20\,Ry and 22\,Ry (20\,Ry corresponding to our standard basis set size). The variations in the determined values for all lateral pair, trio and quattro interactions are within 5\,meV. In contrast, what changes notably is the on-site energy (determined by the $(3 \times 3)$-O structure). With increasing cutoff, its value decreases by more than 100\,meV, and from the convergence pattern using energy cutoffs up to 30\,Ry we expect it to decrease by another $\sim 50$\,meV, when extrapolating to infinite cutoff. The reason for this different convergence behavior is again that the on-site energy gathers all non-coverage dependent contributions. As such, it includes also all the inaccuracies in the description of the isolated O$_2$ molecule, the total energy of which enters the binding energy of all DFT configurations used in the parameterization alike. The slow convergence of the on-site energy is therefore mainly caused by the known slow convergence in the description of the gas phase molecule. At feasible basis set sizes (for full potential calculations of the present system), the lateral interactions can therefore again be determined at a much higher accuracy than the on-site energy.

\begin{table*}
\begin{center}
\caption{\label{tab3}
First-principles lateral interactions obtained using the LDA and the GGA-PBE as exchange-correlation functional for the optimum set of $m=9$ interaction figures determined in Table \ref{tab2}. Additionally shown are the values obtained, when restricting the LGH expansion in eq. (\ref{eqLGH}) to pairwise interactions only. Units are meV.}
\vspace{0.1 cm}
\begin{tabular}{c||c|c|c|c|c||c|c|c|c||c}
Functional &  & \multicolumn{4}{c||}{pair} & \multicolumn{4}{c||}{trio} & quattro \\
  & $E_b^{\rm on-site}$ & $V^{\rm p}_{1}$ & $V^{\rm p}_{2}$ & $V^{\rm p}_{3}$ & $V^{\rm p}_{4}$ & $V^{\rm t}_{1}$ & $V^{\rm t}_{2}$ & $V^{\rm t}_{3}$ & 
$V^{\rm t}_{6}$ & $V^{\rm q}_{2}$ \\ \hline
GGA & 1249 & $-$292 & $-$90 & 50 & 10 &$-$168 & 60 & $-$48 & $-$51 & 120 \\ \hline
LDA & 1968 & $-$344 & $-$128& 54 & 10 &$-$168 & 72 & $-$45 & $-$45 & 96 \\ \hline \hline
GGA, pair only& 1249 & $-$360 & $-$128 & 48 & $-$ & $-$ & $-$ & $-$  & $-$   & $-$ \\
\hline
LDA, pair only& 1968 & $-$410 & $-$160 & 52 & $-$ & $-$ & $-$ & $-$  & $-$   & $-$ \\
\hline
\end{tabular}
\end{center}
\end{table*}

\subsubsection{Approximate exchange-correlation functional}

The uncertainty introduced by the approximate xc functional can not be determined in a comparably quantitative manner as for the numerical approximations, since the exact functional is not known. In order to get at least a feeling for the scatter in the results when using different present-day xc functionals, we evaluated the lateral interactions also within the local density approximation (LDA)\cite{perdew92}, which is a functional that by its very construction is known to yield significantly different adsorbate binding energies than the hitherto employed GGA-PBE functional. Most prominently, the two functionals yield e.g. binding energies for the free O$_2$ molecule that differ by 1.35\,eV (!) when computed with our accurate full-potential LAPW/APW+lo scheme. All 24 DFT input configurations were correspondingly calculated fully relaxed using the LDA and the LDA optimized lattice constant ($a = 3.836$\,{\AA}). Using their energetics in the LGH expansion procedure, we obtain the lateral interaction energies compiled in Table \ref{tab3} for the previously discussed optimum set with $m = 9$ interaction figures.

Comparing to the results obtained before using the GGA-PBE as xc functional, also compiled in Table \ref{tab3}, one clearly observes a similar pattern as before for the numerical uncertainties in the sense that the difference between the two xc functionals shows up predominantly in the on-site energy. In fact, the lateral pair, trio and quattro interactions are strikingly little affected, considering that the two xc functionals yield adsorbate binding energies that differ on the order of $\sim 0.7$\,eV. The reason is again as before, i.e. most of this variation arises out of the description of the free gas phase molecule, which affects all computed configurations alike and therefore solely enters the non-coverage dependent LGH parameter $E_b^{\rm on-site}$. The known larger variation of adsorbate binding energies when using different present-day xc functionals is often put forward as a generic argument against the reliability or usefulness of first-principles lateral interactions. At least for the present system, we can now qualify this concern. Our analysis shows that the on-site energy can indeed not be determined with a high accuracy, cf. Table \ref{tab3}. However, the lateral interactions themselves can be. For the O/Pd(100) system, the uncertainty due to the approximate xc functional seems to be of the order of 50\,meV for the dominant pair interactions. The farther ranging and higher-order interaction terms in Table \ref{tab3} exhibit intriguingly an even smaller scatter and we are currently pursuing further studies to elucidate the generality of this result.

The largely different accuracy with which on-site energy and lateral interactions can be determined allows also to scrutinize when the first-principles parameters may be employed to reliably describe mesoscopic system quantities. We focus here on the effect of the approximate xc functional, since this leads to a larger uncertainty compared to the discussed treatment of the vibrational free energy contributions and the numerical inaccuracies in the total energies. Since the on-site energy has no effect on the order-disorder transition in a canonic adsorbate ensemble, the small variation in the lateral interaction energies exhibited in Table \ref{tab3} rationalizes the good agreement with the experimental critical temperatures. Using the determined LDA interaction values we indeed obtain a $T_c(\Theta)$ curve that is very similar to GGA-PBE as shown in Fig. \ref{fig5}: The overall shape of the curve including the peak structure is completely preserved, and the absolute $T_c$ values are merely shifted by up to 200\,K to higher temperatures. This variation may thus be taken as an indication of the accuracy with which this quantity can presently be calculated from first-principles. For other quantities, where the on-site energy enters explicitly, this accuracy will be significantly worse. Staying with the presently available experimental data for the O/Pd(100) system \cite{chang87}, this would e.g. be the case when aiming to compute the measured temperatures for oxygen desorption, where the on-site energy enters explicitly. 

One can, of course, always argue in terms of a smaller uncertainty by favoring a particular approximate xc functional over others (e.g. because the functional is known to reproduce well similar system quantities). In the sense of first-principles theory, the more appropriate approach would, however, be to eliminate errors by using higher-level theory. In this respect, one could suitably exploit a finding like for the present system that most of the xc uncertainty gets concentrated in the on-site energy by e.g. computing only the latter with a higher-level technique (or appropriate xc-correction scheme \cite{hu07}) and using the already quite accurate lateral interactions determined by present-day xc functionals.

\section{Comparison to empirical parameters}

Instead of a first-principles determination, the traditional method to obtain the strengths of lateral interactions has been to adjust the predictions from atomistic lattice-gas models to experimental observations of the coverage and temperature dependence of on-surface ordering. To keep the number of free fit parameters low, the focus has often been on pair interactions only, i.e. eq. (\ref{eqLGH}) was truncated after the pair terms, and usually also only the dominant short-ranged pair interactions were considered. In this respect, extensive model studies \cite{ree67,binder80,kinzel81,caflisch84,amar84,bak85,bartelt89,liu04} have shown that the $p(2 \times 2)$ and $c(2 \times 2)$ ordering frequently found at (100) cubic metal surfaces requires nearest neighbor repulsions that are so strong that they effectively yield a site exclusion ($-V_1^{\rm p} >> k_BT$), as well as weaker second nearest neighbor repulsions. In order to produce a true $p(2 \times 2)$ ordering further interactions are required. This can either be a third nearest neighbor attraction, or a fourth nearest neighbor repulsion, or more generally a combination of these interactions that fulfills $V_3^{\rm p} - V_4^{\rm p}/2 >0$.\cite{caflisch84,liu04}

Comparing with our first-principles data compiled in Table \ref{tab3}, we see that they perfectly fit into this expected qualitative pattern: Strong first nearest neighbor repulsion, weaker second nearest neighbor repulsion, and the combination $V_3^{\rm p} - V_4^{\rm p}/2$ is attractive. However, differences arise when turning to a more detailed modeling of the experimental O/Pd(100) phase diagram of Chang and Thiel \cite{chang87}. For a lattice-gas model considering nearest-neighbor site exclusion and finite second and third nearest neighbor interactions, Liu and Evans showed that the best overall topological agreement with the experimental phase diagram is achieved for $V_3^{\rm p} \approx -V_2^{\rm p}$.\cite{liu04} In particular, the position of the disordered to $c(2 \times 2)$-O transition line shifts notably with the ratio of $-V_2^{\rm p}/V_3^{\rm p}$, and for a ratio of $\sim 1$ the $T_c(\Theta)$ curve no longer exhibits the peak structure at $\Theta =0.25$\,ML visible in the first-principles curve in Fig. \ref{fig5}. Since this resembles the shape of the experimental curve better, it could indicate that the calculated first-principles ratio of $-V_2^{\rm p}/V_3^{\rm p} \approx 3$ is too large. On the other hand, it is intriguing to see that both LDA and GGA-PBE yield roughly the same $-V_2^{\rm p}/V_3^{\rm p}$ ratio and in turn clearly a peak in the $T_c(\Theta)$ curve, cf. Fig. \ref{fig5}. Since Chang and Thiel discuss also a rather large uncertainty of $\sim 0.05$\,ML in their experimental coverage calibration \cite{chang87}, a careful remeasurement of the experimental $T_c(\Theta)$ curve would thus be very interesting to fully settle this point.

A much more fundamental difference between the set of empirical and first-principles lateral interactions is that the former is restricted to just pairwise interactions. To test the validity of this frequently applied approximation, we restrict our LGH expansion to pair interactions up to third nearest neighbors and repeat the parameterization using exactly the same 24 DFT configurations as in Section IIIB. The resulting interactions are also compiled in Table \ref{tab3}, using the LDA or GGA-PBE as xc functional. Comparing with the values obtained for each functional from the proper LGH expansion (and using exactly the same first-principles input), we note significant differences. The dominant short-ranged interactions are overestimated, in the case of the first nearest neighbor pair interaction by $\sim 60$\,meV. Overall, the lateral interactions are too repulsive, which is a consequence of them having to mimic the effect of the neglected overall repulsive trio interactions. As expected, the true microscopic lateral interactions are thus blurred by spurious contributions arising from the invalid assumption that lateral interactions at metal surfaces could be expressible in pair-only terms. The invalidity of this assumption is already clearly demonstrated by the similar magnitude of the dominant trio interactions ($V^{\rm t}_1, V^{\rm t}_2$, etc. in Table \ref{tab3}) compared to the dominant short-ranged pair interactions, which necessarily reduces pair-only restricted expansions to a set of {\em effective} parameters at best.

\begin{figure}
\scalebox{0.40}{\includegraphics{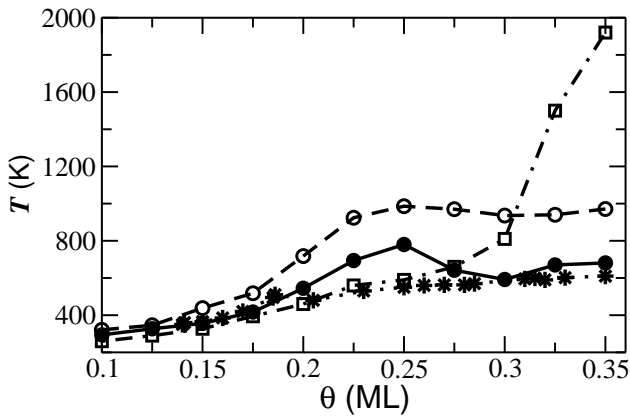}}
\caption{\label{fig7}
$\Theta-T$ diagram of the critical temperatures $T_c$ for the order-disorder transition. Shown with stars are the experimental data from Ref. \onlinecite{chang87}, and with solid circles the values obtained when using the optimum GGA-based LGH expansion, cf. Table \ref{tab3}. These curves are compared to those obtained, when considering only pair interactions in the LGH: Using the GGA values compiled in Table \ref{tab3} (empty circles), or the empirically adjusted values by Liu and Evans (empty squares), see text.}
\end{figure}

The deficiencies introduced by such an effective description are not only the lack of microscopic meaning of the lateral interactions themselves, but also errors in the mesoscopic system properties calculated with these interactions. This is again exemplified for the order-disorder transition in Fig. \ref{fig7}, which shows that the $T_c(\Theta)$ curve derived from the pair-only lateral interactions in Table \ref{tab3} differs notably from the curve derived from the corresponding proper LGH expanded lateral interactions. The critical temperatures are up to 350\,K higher, even though being based on exactly the same first-principles energetics. The shortcomings due to the restricted LGH can also not be overcome, when turning to a completely empirical approach and adjusting the lateral interaction values to reproduce experiment. Within the above described lattice-gas model with pair interactions to third nearest neighbor and fitting to the critical temperatures in the coverage range below 0.25\,ML, Liu and Evans determined $V_3^{\rm p} = -V_2^{\rm p} \approx 40$\,meV \cite{liu04}, i.e. lateral interactions that differ considerably from the first-principles values compiled in Table \ref{tab3}. As apparent from Fig. \ref{fig7}, these empirical parameters do indeed lead to an excellent fit of the experimental $T_c(\Theta)$ curve in the coverage range up to 0.25\,ML. However, outside the fitted coverage range, the effective description rapidly breaks down and predicts a critical temperature curve in qualitative disagreement with experiment. Already at a coverage of 0.35\,ML, i.e. only 0.1\,ML outside the fitted coverage range, the predicted critical temperature for the order-disorder transition is wrong by 1200\,K. 

The spurious contributions contained in the effective pair interactions also prohibit a systematic determination of which pairwise interactions to consider. In the systematic LGH expansion detailed in Section IIIB, the lateral interactions are chosen out of a larger pool of lateral interactions and the ultimately determined pair interactions decrease in magnitude with distance. In contrast, when we extend the pair-only expansion to include the fourth and fifth nearest neighbor interaction, we obtain a negligible $V_4^{\rm p}$ of -2\,meV, but a $V_5^{\rm p}$ of -18\,meV. The spurious contributions contained in the effective $V_5^{\rm p}$ thus feign that such far ranging interactions are significant, whereas the proper first-principles LGH expansion shows no significant interactions between adsorbates at distances larger than the fourth nearest neighbor site. In empirical approaches this uncertainty (or degree of arbitrariness) as to which interactions to consider is even further aggravated, since there are typically several sets of pairwise interactions which equally well fit the less stringent mesoscopic ordering behavior.

\section{Conclusions}

In summary, we have presented a first-principles lattice-gas Hamiltonian study of the on-surface ordering behavior of O adatoms at the Pd(100) surface. A key feature of the approach is the systematics of the LGH expansion, both with respect to the selection of the considered lateral interactions and with respect to the ordered configurations, the DFT energetics of which is employed in the parameterization. In contrast to empirical or semi-empirical {\em ad hoc} parameterizations the approach does provide an error-controlled link between the electronic structure regime and the system's mesoscopic ensemble properties. Carefully scrutinizing the approximations entering at the level of the underlying DFT energetics, at the LGH interface and in the statistical simulations, we identify the approximate exchange-correlation functional employed to obtain the DFT energetics as the major source of uncertainty in the approach. This uncertainty affects predominantly the on-site energy, and only to a much lesser degree the lateral interactions themselves. Comparing LDA and GGA exchange-correlation functionals, we estimate the accuracy of the latter to be within $\sim 60$\,meV for the present system. This shows that the known much larger variation of adsorbate binding energies with different xc functionals can not be used to uncritically question the reliability or usefulness of lateral interactions derived from first-principles {\em per se}.

The rather high accuracy of the first-principles lateral interactions also rationalizes the obtained good agreement with the experimental low coverage phase diagram for the O/Pd(100) system, i.e. the ground state ordered structures and critical temperatures for the order-disorder transition. An important feature of the set of lateral interactions is that it contains many-body interactions of comparable magnitude to the dominant short-ranged pair interactions. This demonstrates the invalidity of the assumption of exclusively pairwise interactions between adsorbates at metal surfaces that is frequently applied in empirical approaches. Indeed, when restricting the LGH expansion to just pairwise lateral interactions, we find that this results in effective interactions which contain spurious contributions that are of equal size, if not larger than any of the uncertainties e.g. due to the approximate xc functional. These effective parameters lack microscopic meaning and lead to uncontrolled errors in the mesoscopic system properties calculated with them. In the present system, this is illustrated clearly by critical temperatures that are up to 350\,K higher than the ones obtained with the proper set of lateral interactions, even though both sets were based on exactly the same first-principles energetics. 

We conclude that at least for the present system, an appropriate first-principles statistical mechanics framework and present-day DFT energetics can determine lateral interactions at least as reliably as the traditional empirical approaches, and without suffering from the non-uniqueness in the selection of pairwise-additive adspecies interactions which reasonably fit available data.

\section*{Acknowledgements}

The EU is acknowledged for financial support under contract no. NMP3-CT-2003-505670 (NanO$_2$). We wish to thank John Kitchin, Cesar Lazo, Jutta Rogal and Matthias Scheffler for stimulating discussions.


\begin{thebibliography}{99}
\bibitem{footnote}
An alternative formulation equally found in the literature would be to let all sums run independently over the lattice sites, necessarily implying a multiple counting of site contributions. Pair interactions would then e.g. be double counted, which needs to be taken into account, when comparing interaction values quoted within this formulation to values quoted within the single-counting formulation employed here.
\end{thebibliography}
\end{document}